\documentclass[runningheads]{llncs}
\usepackage{graphicx}
\usepackage[frozencache=true]{minted}
\usepackage{hhline}

\usepackage{url}
\usepackage{array}

\begin{document}

	\title{An AST-based Code Change Representation and its Performance in Just-in-time Vulnerability Prediction}
	\titlerunning{A Code Change Representation for Just-in-time Vulnerability Prediction}
	\author{Tamás Aladics\inst{1,2}\orcidID{0000-0002-4689-8878} \and
		Péter Hegedűs\inst{1,2}\orcidID{0000-0003-4592-6504} \and
		Rudolf Ferenc\inst{1}\orcidID{0000-0001-8897-7403}}
	\authorrunning{T. Aladics et al.}
	\institute{University of Szeged, Szeged, Hungary \and
		FrontEndART Ltd., Szeged, Hungary \\
		\email{\{aladics,hpeter,ferenc\}@inf.u-szeged.hu}\\
	}
	\maketitle              % typeset the header of the contribution

	\begin{abstract}
		The presence of software vulnerabilities is an ever-growing issue in software development. In most cases, it is desirable to detect vulnerabilities as early as possible, preferably in a just-in-time manner, when the vulnerable piece is added to the code base. The industry has a hard time combating this problem as manual inspection is costly and traditional means, such as rule-based bug detection, are not robust enough to follow the pace of the emergence of new vulnerabilities. The actively researched field of machine learning could help in such situations as models can be trained to detect vulnerable patterns. However, machine learning models work well only if the data is appropriately represented. In our work, we propose a novel way of representing changes in source code (i.e. code commits), the Code Change Tree, a form that is designed to keep only the differences between two abstract syntax trees of Java source code. We compared its effectiveness in predicting if a code change introduces a vulnerability against multiple representation types and evaluated them by a number of machine learning models as a baseline. The evaluation is done on a novel dataset that we published as part of our contributions using a 2-phase dataset generator method. Based on our evaluation we concluded that using Code Change Tree is a valid and effective choice to represent source code changes as it improves performance.
		
	\end{abstract}
	\section{Introduction}
	
	Security is an important aspect of software development and as technology progresses it seems to become increasingly hard to handle the rapid increase in vulnerabilities. According to the Mend database which aggregates data from NVD and other vulnerability databases, the number of published open source vulnerabilities is on a steep rise as in 2020 this number has increased by 50\%. \cite{johnson_2022}
	
	In software security, finding vulnerabilities as soon as possible is important to minimize the possibility of harm done and to reduce the additional cost of fixing the security holes later. The earliest point when vulnerabilities could be reliably identified in the development pipeline is the time when they are added to the code base, that is, when the changes are added to the version control system (i.e. commit time). Another positive effect of this is that the developers receive immediate feedback regarding possible issues in the software. This practice is called just-in-time vulnerability prediction \cite{KAMEI}.
	
	Researchers and interested parties in the industry are actively trying to find ways to become more resilient to software vulnerabilities. The most prominent and sustainable way of this is to use automated techniques to detect if a piece of software is likely to be vulnerable \cite{MCKINNEL2019175}\cite{hydara}.
	These approaches usually involve extensive static analysis based on rule sets and manually developed pattern matchers, or dynamic analysis based on executing the software. Numerous tools exist for this purpose such as SonarQube  \footnote{https://www.sonarqube.org} and the Clang Sanitizers \footnote{https://github.com/google/sanitizers}. Companies have widely adopted these tools to improve the quality of their product and to become less prone to vulnerabilities.
	
	Even though these techniques definitely improve the quality of software, from the point of vulnerability and bug detection they are reported to be mostly ineffective in practice as they can't adapt fast enough to the new vulnerabilities, have scalability issues and/or have high false positive rates \cite{johnson}\cite{antunes}\cite{MORRISON}.
	
	The trending field of machine learning has proven to successfully solve problems of similar type, when there is an abundance of data yet the problem is hard to formulate or manual, human-powered solutions are too resource-demanding. However, a ML model can only be as good as the data it is trained on. In the case of source code, the form of representation is not trivial: it can be used as raw text, intermediate representations such as abstract syntax trees or control flow graphs can be used or indirect attributes are possible forms too, like software quality metrics \cite{ZHANG} \cite{HARER}.

	Even though there has been substantial research done in just-in-time (JIT) vulnerability prediction, current techniques have little potential in practical usage as their results are usually not localized, have low recall/precision, or are not reproducible \cite{NEUHAUS}\cite{LIVHITS}\cite{PERL}. One of the main challenges that make just-in-time vulnerability prediction hard is that automatization involves finding appropriate ways to represent differences between two states of software: the pre-commit and post-commit states. In a recent analysis of the state of JIT vulnerability prediction, it is shown that using existing metrics and textual features to represent changes in software is not sufficient enough \cite{LOMIO}. As our attempt to remedy this situation, we focused on representing this change as differences in source codes as a commit's most vital information is the actual source code it encapsulates. 
	
	In our work, we show multiple ways and also propose a novel method for code change representation, the Code Change Tree. To showcase these methods we generate these representations for every entry of a lately released vulnerability introducing commit (VIC) dataset. This generation process is not trivial as VIC databases are scarce in number and can not be directly mapped from existing vulnerability fixing commit (VFC) databases, such as NVD. Therefore, we used our SZZ-based approach\cite{VULNDB} which has two phases: in the first phase it generates candidate commits using SZZ, and in the second phase it filters the candidate commits based on scores referenced as relevance scores.
	
	Then, we train multiple machine learning models on these representations and compare their predictive power. Our results are interpreted through the following research questions:
	
	\begin{description}
		
		\item[RQ. 1] Can a vulnerability introducing database generated from a vulnerability fixing commit database be used for vulnerability prediction? 
		\item[RQ. 2] How effective are Code Change Trees in representing source code changes?
		\item[RQ. 3] Are source code metrics sufficient to represent code changes?
		
	\end{description}
	
	This work is an extension of our previous conference paper\cite{VULNDB} adding the following new contributions:
	\begin{itemize}
		\item A novel way of representing source code changes, Code Change Tree;
		\item Comparison of multiple code change representation forms;
		\item Evaluation of the proposed code change representation for just-in-time vulnerability prediction based on the data published in the conference paper \cite{VULNDB}.
	\end{itemize}

	\section{Related work}
	
	Heavily motivated by the industry and software security, vulnerability prediction is an actively researched field. The main line of traditional vulnerability prediction models (VPMs) is based on using software metrics - quality indicators derived from the software. 
	
	Shin et al. used VPMs using code complexity metrics,  code churn, and fault history as predictors on the code base of Mozilla Firefox web browser \cite{SHIN}. They found that fault prediction models can be used for specific cases of vulnerabilities, however, both fault prediction and vulnerability prediction models require significant improvement to reduce false positives while providing high recall. Similarly, other works use LOC metrics supplemented with complexity metrics to train neural networks \cite{ZAGANE} or mine features based on text mining techniques as input for VPMs \cite{Scandariato2014PredictingVS}.
	
	As a vulnerability's severity is directly influenced by how long it remains unexposed, finding vulnerabilities as soon as possible has become a key endeavor for software security research communities. Recently VPMs are actively getting developed to predict vulnerabilities when they are added to the version control system, that is, in commit time. These methods can use different parts of commits as predictors, such as the commit messages and bug reports\cite{ZHU} or the source code's before and after states \cite{MINH}. A commonly referenced work in this topic is VCCFinder, a tool that combines code metrics with GitHub metadata for features and trains an SVM model on them \cite{VCCFinder}\cite{Riom2021RevisitingTV}.
	
	As reported by Lomio et al. existing metrics are not sufficient enough \cite{LOMIO}. In their research, they used several different code metrics and textual features derived from the bag of words representations of the before and after commit states, similarly to VCCFinder. Our work's aim is to contribute to the solution of this problem and provide a way to represent source code changes. 
	
	To gain insight it is beneficial to look at related research in source code representation. Apart from metrics, which have already been discussed as a possible form of software state representation, other approaches can use intermediate forms of the source code as input. DeFreez et al. use control flow graphs and perform random walks on inter-procedural paths in the program and use these paths as features \cite{DEFREEZ}. Devlin et al. embed functions by doing a depth-first traversal on the corresponding abstract syntax tree (AST)\cite{DEVLIN}, while Pan et al. used a convolutional neural network architecture to extract features from ASTs \cite{PAN}. Gaining inspiration from these and other works, we choose to use ASTs to represent code changes.
	
	However, since code changes include a before and after state, each code change has two ASTs corresponding to them: the before and after state's AST. Since we are interested in the difference, or change between the two ASTs, we propose an AST-like structure that we call Code Change Tree, which incorporates the differences between the two ASTs. 
	
	After getting the Code Change Tree for a commit, a way of mapping the tree to a vector is needed, so that later it can be fed into machine learning models as inputs. We followed the same approach as we did in our previous work \cite{ALADICS}, where we performed depth-first traversal on the AST and used Doc2Vec embedding \cite{DOC2VEC} to get fixed size vectors that keep structural information \cite{ALADICS}. We choose this method as it is fast to compute and has low resource needs while also having solid theoretical background based on Word2Vec.
	
	Regarding change representation, lately various approaches have been published, one of the most prominent is CC2Vec\cite{CC2VEC}. CC2Vec uses a hierarchical attention network to extract features from commits. This work is promising, however, it has some differences compared to our method: Code Change Trees represent changes from any two code piece that has corresponding AST, so it is not restricted to commits as CC2Vec (which uses commit-specific metadata, such as commit messages). Also, CC2Vec represents the differences in code changes as the added/removed lines as tokens, while in our work we aim to keep the structural information that ASTs have by representing changes in a tree structure.
	
	Another work in this topic is Commit2Vec \cite{commit2vec}, which uses a structure derived from Code2Vec that leverages path-contexts: paths in the AST from leaf node to leaf node. We couldn't use this approach as the tools and the code wasn't published. Also, its methodology is different as it learns on the extracted paths, while our approach constructs a tree to keep the structural attributes of the changes.

	\section{Methdology}
	\label{section:methodology}
	
	\subsection{Overview}
	\label{section:overview}
	
	\begin{figure}[htb!]
		
	\begin{minted}[linenos]{Java}
	//Before state
	class HelloWorld {
		public static void main(String[] args) {
			System.out.println("Hello, World!"); 
		}
	}
	\end{minted}
	\hrulefill
	\begin{minted}[linenos]{Java}
	//After state
	class HelloWorld {
		public static void main(String[] args) {
			String msg = "World!";
			System.out.println("Hello, World!");
			System.out.println("Hello, " + msg );
		}
	}
	\end{minted}
	\caption{An example source code change}
	\label{fig:change_example}
	\end{figure} 
	
	As mentioned in the previous sections of our work we try to provide a novel way of presenting source code changes: differences between two states of source code. More specifically, let $S = \{t_1, t_2, ..., t_n\}$ be a state of a source code, where $t_i$ are tokens that are present in the source code. Then, a code change is a pair of states that are consequent in time, namely the state before the change $S_{pre} = \{t_1, t_2, ..., t_n\}$, and $S_{post} = \{t_1, t_2, ..., t_m\}$, the state after the change. A simple code change example can be seen in Figure \ref{fig:change_example}.
	
	Rather than only considering changes in code as raw text (i.e. changes in the token sequences $S_{pre}$ and $S_{post}$), we try to represent changes happening at a structural level. A good way to reason about the structure is to use an intermediate representation, such as the abstract syntax tree (AST), which can be computed directly from the source code, that is, for each source code state $S$ there exists a corresponding AST (a more formal introduction to ASTs can be found in Section \ref{section:simple_code_change}). One of our main contributions in this work is to represent the differences between $AST_{pre}$ and $AST_{post}$ in a tree.
	
	Also, when reasoning about software representations, another fundamental question is the localization of the prediction, at what granularity is the source code processed. Usually, it happens at statement, method, class, or file level. Since we decided to use the AST as the base structure, any source code element can be represented by our method that has a corresponding AST. In this report, we decided to use function-level predictions as it is more specific than class-level predictions but still has a corresponding AST with meaningful structure -unlike statements, where the corresponding AST is usually too shallow.
	
	To showcase our approach we describe three different approaches to represent changes in source code: a source code metric-based approach (Section \ref{section:metric_based}), a simple code change approach (Section \ref{section:simple_code_change}), which is based on concatenating the flattened ASTs of the before and after states, and the Code Change Tree approach (Section \ref{section:code_change_tree}), which is based on concatenating the flattened Code Change Trees of the before and after states. 
	
	However, in the case of simple code change and Code Change Tree approaches the flattened before and after states are sequences of tokens which is not a valid input for ML models. To remedy this situation, we employ an embedding from sequences of tokens to vectors using Doc2Vec (Section \ref{section:doc2vec}).

	\subsection{Metric based approach}
	\label{section:metric_based}
	Many works in source code representation use derived features from the code as predictors, such as source code metrics \cite{NGUYEN}\cite{SHIN_EVAL}. Source code metrics can be good indicators of many of the software's attributes, such as lines of code (LOC), nesting level(NL) and McCabe’s cyclomatic complexity (MCC) \cite{MCCABE}. 
	
	As one of our baselines, we choose to represent changes by aggregating metrics calculated for the before and after state of a code piece. In detail, for a function that is part of a software change's before/after state we measured 37 metrics using the open-source static analysis tool SourceMeter.\footnote{https://sourcemeter.com/} Then, concatenated the metrics corresponding to the before and after states ($S_{pre}$ and $S_{post}$) of the function to form a change representation. This way, the source code change in a function consisting of $S_{pre}$ and $S_{post}$ are represented as a vector of 74 elements.

	\subsection{Simple code change approach} 
	\label{section:simple_code_change}
	To reflect on the structure of source code, an intermediate representation form (mainly used by compilers), the abstract syntax tree (AST) is considered to be the preferred form in many techniques \cite{CODE2VEC}\cite{CTX_AWARE}\cite{commit2vec} as it stores syntactical features on its edges and in the node attributes. 
	
	As the AST is an important concept in our approach, we provide a formal definition.
	\begin{definition}[AST]
		An abstract syntax tree is a structure of form (N,L,T,r, $\Omega$, $\Psi$), where:
		\begin{itemize}
			\item N is the set of non-terminals (non-leaf nodes)
			\item L is the set of terminals (leaf nodes)
			\item T is a set of tokens
			\item r is the root node
			\item $\Omega$ is a mapping from nodes to their children: $N \Rightarrow N \cup L  $
			\item $\Psi$ is a mapping from terminals to their corresponding code tokens: $L \Rightarrow T$
		\end{itemize}
	\end{definition}
	
	An advantage of AST is that it can be directly calculated for source code elements without the need to execute the software, and as such, for each we can generate the corresponding $AST_{pre}$ $AST_{post}$ representations for the source code change states $S_{pre}$ and $S_{post}$.

	Unfortunately, most machine learning models cannot take tree structures as input, so using an AST directly is not suitable. To make an AST usable as input, we need a way to map it to a sequence of tokens. This process is called flattening and there are multiple ways to do it. In our work, we choose to flatten a tree structure by traversing it in a depth-first manner and we represent each node $n \in N \cup L$ in the tree, based on whether it's a terminal or a non-terminal:
	\begin{itemize}
		\item if $n \in N$ return the type of n
		\item if $n \in L$ return the type and value of n
	\end{itemize}
	Note that this way of representing nodes is analogous to many works in the area, where they decided to keep the terminal's value, as it holds important semantic information \cite{CODE2VEC}\cite{CODE2SEQ}. 
	
	To provide a baseline for AST-based techniques, we flattened $AST_{pre}$ and $AST_{post}$ using the method we just introduced to get sequences of node representations that can be used in ML models. However, these sequences can be of variable length and content, so we employ an embedding technique based on Doc2Vec to generate fixed-length vectors. For further details regarding the sequence embedding please refer to Section \ref{section:doc2vec}.

	\subsection{Code Change Tree approach}
	\label{section:code_change_tree}
	The representation method introduced in Section \ref{section:simple_code_change} does inherently capture syntactical information, as it uses the flattened AST for the before and after states. However, it is sub-optimal in representing changes as it contains the whole function (or any other AST structure, depending on granularity as discussed in Section \ref{section:overview}). For example, in Figure \ref{fig:change_example}, the flattening considers every element of the code, even those that are unchanged, such as the class, function definition, and the line that calls \mintinline{Java}|System.out.println| with the \mintinline{Java}|"Hello, World!"| string as parameter.
	
	As a way to represent only the changes, we designed a novel structure, Code Change Tree, that aims to capture only the differences between two ASTs: it can be calculated for an AST (reference AST) to represent changes to another AST (target AST). To find the changed parts we represent each tree as a set of unique paths from their root to each terminal. Then, the root paths that are identical in both trees are discarded from the reference AST's set of root paths. Finally, a tree constructed from the reference AST's root paths, which we refer to as Code Change Tree. To give more insight into our method, in the following we provide a more precise description of our method and its components.
	
	To reason about specific parts of an AST, a more fine-grained processing is needed than flattening, because as reported by Alon et al. \cite{STRUCTURAL_MODELS}, this method creates artificially long distances between the node and ancestor nodes. A solution to this is to consider an AST as a set of unique paths, which is referred to as path-based representation \cite{PATH_BASED_REPR}. We decided to use a root-path based representation. Formally:
	\begin{definition}[AST Path]
		An AST path $p$ of length $k$ is a sequence $n_1, n_2, \dots,$ $ n_{k+1}, n_i \in N \cup L $, where $n_{i+1} \in \Omega(n_i)$ (that is, each element is a child of the preceding element). For convenience we also define $start(p)$ as $n_1$ and $end(p)$ as $n_{k+1}$
	\end{definition}
	
	\begin{definition}[AST root-path]
		An AST root-path $rpath$ is an AST path, where $start(rpath) = r$ and $end(rpath) \in L$ 
	\end{definition}
	
	Using root-paths, we can represent an AST by generating the root-path for each terminal or formally:
	
	\begin{definition}[Root-path based AST representation]
		A root-path based representation can be defined for an AST as a set of unique root-paths $rpath_1,$ $rpath_2, \dots rpath_n$ where $end(rpath_i) \in L$ and $\forall n \in L$ exists an $rpath_i$ such that $rpath_i = n$.
	\end{definition}
	
	As each AST can be represented as a set of root-paths, defining a difference between them is needed. To explain the representation of differences between two ASTs, let $AST_{ref}$ be the reference AST (in which we are interested in the changes) and a $AST_{target}$ be the target AST (to which we compare $AST_{ref}$). 
	To generate the Code Change Tree for $AST_{ref}$ with respect to $AST_{target}$ the root-path based representations must be calculated for both $AST_{ref}$ and $AST_{target}$, and discard the identical root-paths from the set of root-paths corresponding to $AST_{ref}$. More specifically:
	
	\begin{definition}[Root-path based AST difference]
		Let $rpaths_{ref}$ and\\$rpaths_{target}$ be the root-path based AST representations for $AST_{ref}$ and $AST_{target}$ respectively. Note that both $rpaths_{ref}$ and $rpaths_{target}$ are sets.
		
		The root-path based AST difference for $AST_{ref}$ with respect to $AST_{target}$ can be calculated as $rpaths_{ref}$ $\setminus$ $rpaths_{target} $
	\end{definition}
	
	However, to make $rpaths_{ref}$ $\setminus$ $rpaths_{target} $ a valid operation, the equality of two root-paths must be defined, so that identical root-paths can be removed from $rpaths_{ref}$. Straightforwardly, we define that two root-paths are equal if they have the same length and at every position the nodes are equal.
	
	\begin{definition}[Root-path equality]
		Let $a = a_1, a_2, \dots a_n$ and $b = b_1, b_2, \dots b_m$ be root-paths. $a$ represents the same root-path as $b$, if $n = m$ and $id(a_i) = id(b_i) \> \forall i \in 1, \dots, n$, where $id$ is a mapping from an AST node to an identifier
		
	\end{definition}
	
	Note that root-paths equality depends on the nodes' equality, which is determined by the $id$ mapping that assigns an identifier to a node in an AST. This way, root-path can be compared in a cross-AST manner, as nodes between different ASTs can be compared. This means that the $id$ mapping is a fundamental decision in Code Change Trees. In our work, we propose an $id$ mapping we designed empirically by trial and error, but we don't exclude the possibility that better solutions exist. Our id mapping is a recursive one, that concatenates the ids of every ancestor with the node's child rank (position related to siblings: the leftmost children has child rank 0, etc.) and the node's type. This mapping is more precisely defined through the pseudo-code in \ref{figure:pseudo_id}.
	
	\begin{figure}
		\begin{minted}[tabsize=4, linenos]{Python}
			def id(node):
				ancestors_id = ""
				for ancestor in node.ancestors:
					ancestors_id += id(ancestor)
				
				identifier = ancestors_id + node.child_rank + node.type
		\end{minted}
		\caption{Pseudo code of our proposed id mapping}
		\label{figure:pseudo_id}
	\end{figure}
	
	This mapping has the advantage that it not only encapsulates local attributes about the node but some contextual information as it concatenates the ancestors' ids too. This way, a node can only be equal to another node if the path leading to it from the root node is the same.
	
	Finally, with the definition of root-path differences, a tree representing the differences between $AST_{ref}$ and $AST_{target}$ can easily be constructed from the root-paths in $rpaths_{ref}$ (after calculating the root-path difference with $rpaths_{target}$), by creating an empty tree and extending it while iterating through each root-path. We provide implementation details with the help of the pseudo-code in Figure \ref{figure:pseudo_tree}.
	
	\begin{figure}
		\begin{minted}[tabsize=4, linenos]{Python}
			def construct_change_tree(root_paths):
				tree = ChangeTree()
				for rpath in root_paths:
					current_change_tree_node = tree.root 
				for node in rpath:
					extend_tree = True
				for child in current_change_tree_node.children:
					if id(child) == id(node):
					current_change_tree_node = child
					extend_tree = False
				
				if extend_tree:
					current_change_tree_node.children.add(node)
			
		\end{minted}
		\caption{Pseudo code of the tree construction from a set of root paths}
		\label{figure:pseudo_tree}
	\end{figure}
	
	With Code Change Tree defined, it is possible to describe our method of representing source code changes. As introduced in Section \ref{section:overview}, a code change is a pair of states $S_{pre}$ and $S_{post}$. Then, the corresponding ASTs to the states ($AST_{pre}$ and $AST_{post}$) respectively can be generated. In our work, we use the TreeSitter \footnote{https://tree-sitter.github.io/tree-sitter/} tool to generate the ASTs. Employing the steps we introduced before in this section, we generate two Code Change Trees for the before and after states, containing the changes relative to each other. Using the root-path notation, we take $AST_{pre}$ and $AST_{post}$, generate root-path based representations for them and calculate the root-path differences with $AST_{pre}$ as a reference and $AST_{post}$ as target and vice-versa. Based on the root-path differences, the Code Change Trees for $S_{pre}$ and $S_{post}$ can be calculated.
	
	\subsubsection{Example} 
	\begin{figure}
		\includegraphics[width=\textwidth]{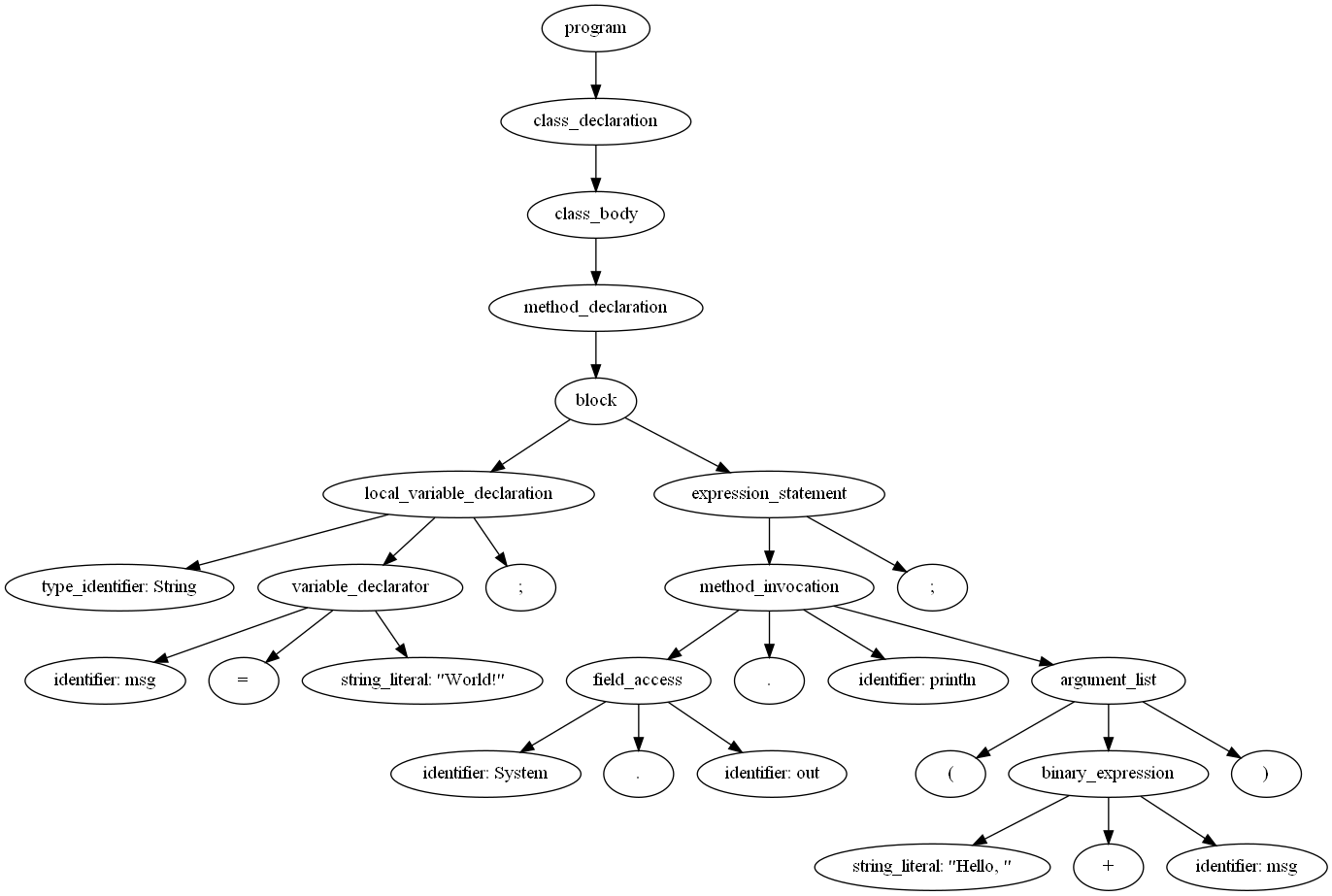}
		\caption{Code Change Tree for Figure \ref{fig:change_example}}
		\label{fig:ch_tree_example}
	\end{figure}
	
	To demonstrate the usage of Code Change Trees to represent changes, we prepared a simple case study based on Figure \ref{fig:change_example}. Here, the before state ($S_{pre}$) consists of a minimal program with a function call, the traditional "Hello World" printout. The change is adding a line before, and after the function call ($S_{post}$). This change offsets the original method call's position and also adds additional lines so this modification is not trivial to detect. Normally, we would have to calculate the Code Change Trees for both $S_{pre}$ and $S_post$, in this case however the root-path differences for $S_{pre}$ will result in an empty set (every path in $S_{pre}$ is present in $S_{post}$), so only the root-path differences for $S_{post}$ are relevant. The Code Change Tree generated from the latter can be seen on Figure \ref{fig:ch_tree_example}.
	
	To compare these results with the simple code change representation shown in Section \ref{section:simple_code_change}, we generated the ASTs (by using TreeSitter) and flattened them. The before-change state for the simple code change representation contained 31 nodes, the after-change state 55. Using Code Change Tree, the before state has 0 tokens (as it was an empty tree since no paths are unique in it compared to the after state), and the after state has 28 tokens. Overall, representing both states for the simple code change representation would take 86 tokens, while Code Change Tree would take only 28. This is a substantial reduction that keeps relevant data, including contextual information, as whole paths corresponding to a change are stored.  
	
	\subsection{Doc2Vec embedding}
	\label{section:doc2vec}
	In Section \ref{section:metric_based} we introduced the usage of metrics for code change representation, which generates numerical values as features, thus they can be directly fed into machine learning models as input. Unfortunately, this is not the case for simple code change (Section \ref{section:simple_code_change}) and Code Change Tree (\ref{section:code_change_tree}) based representations, as in both cases the outputs are two trees: ASTs and Code Change Trees corresponding to the before and after change states. These trees are then flattened as described in Section \ref{section:simple_code_change} to sequences of tokens, which are still not numeric values.
	
	To make the tree-based methods applicable to ML models a way to map the sequences to numeric input is needed. We use the technique from our previous work \cite{ALADICS} that uses Doc2Vec with some improvements, to embed the token sequences to fixed length vectors. 
	In more detail, to start we selected 2 million methods randomly from the GitHub Java Corpus \cite{githubCorpus2013}. We flattened these methods to sequences of tokens, then we preprocessed each of them in line with common NLP practices using the Genism library \cite{rehurek2011gensim}: replaced whitespaces in strings with underscores and replaced tokens which were not present in at least 1\% of the sequences (20.000 methods) with an OOV (out of vocabulary) constant. We trained the Doc2Vec model on this corpus.
	
	Training this model results in a Doc2Vec model that was tailored for Java methods by the design of distributed representations: the learned embedding process generates vectors in a vector space where vectors corresponding to similar methods have ideally small distances. With such a model ready, we used it to embed the variable length sequences to fixed length vectors in Section \ref{section:simple_code_change} and \ref{section:code_change_tree}, as the last steps before evaluation in Section \ref{section:evaluation}.
	
	\subsection{Vulnerability DB}
	\label{section:vuln_db}
	
	\begin{figure}
		\includegraphics[width=\textwidth]{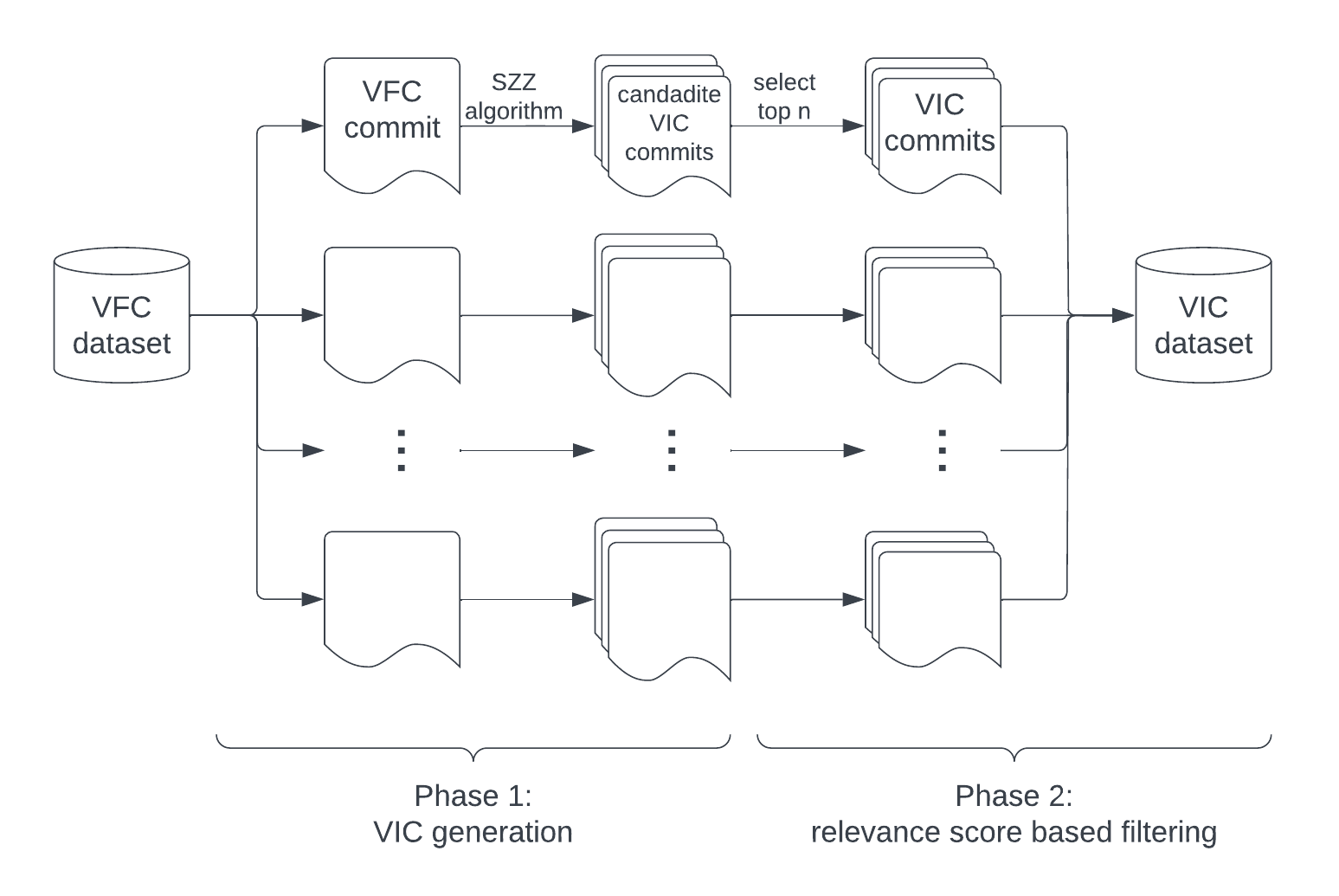}
		\caption{Overview of the VIC dataset generation process}
		\label{figure:vuln_db}
	\end{figure}
	
	This work regarding Code Change Trees is a natural extension to the work we previously published \cite{VULNDB} with a vulnerability introducing commit (VIC) database generation process and an actual dataset generated by following our approach. As such, we used this dataset for evaluation purposes. The main idea of this approach is that while there is an abundance of vulnerability fixing commit (VFC) datasets, vulnerability introducing commit (VIC) datasets are in short supply. As a solution, we propose a method that generates VIC datasets from VFC datasets.
	
	The generation process has two phases. The first is the preliminary VIC candidate detection, which takes a VFC database as input, and outputs a set of VIC for each VFC using the SZZ algorithm \cite{szz}. However, SZZ is reported to output false positives in many cases and a variable number of results without any order or explanation. The second phase of our algorithm is designed to improve on these weaknesses by performing a filtering based on a score calculated for each candidate commit, referred to as the relevance score. Based on the relevance score we can provide an ordering for the candidate commits and also select the top n best fit. This overview of this process can be observed in Figure \ref{figure:vuln_db}
	
	As part of the work in \cite{VULNDB}, we applied our algorithm with the input VFC dataset being the project-KB \cite{ponta2019msr}, a dataset that is manually curated for open source Java projects, with each entry having a reference to the US Nation Vulnerability Database \cite{NVD}. We used a recent implementation of the SZZ for the first phase, SZZUnleashed \cite{Borg_2019} and for the second phase, we selected the top 2 best fit VIC for each VFC. The resulting dataset has 564 VFC entries in 198 open-source projects with each of them having at most two VIC commits but on average at least one.

	\section{Results}\label{section:evaluation}
	
	To evaluate the approaches we introduced in Section \ref{section:methodology}, we aimed to reason about vulnerability detection at the function level. For that purpose mined a number of vulnerable and non-vulnerable methods by traversing the VIC dataset described in Section \ref{section:vuln_db}, and for each VIC we looked for functions that were changed in Java files. Similarly, for the VFC entries, we extract the functions that were changed and define the changes' before states with the functions from the VIC and the after states with the functions from the VFC. For non-vulnerability-related function changes, we randomly picked functions from the VIC commit that have no corresponding function in the VFC (this is because we assume that only files changed in the fixing commit were vulnerable). This way, we produced 5934 vulnerable and 29.670 non-vulnerable methods, so that the ratio of vulnerable to non-vulnerable entries is 20\% and consequently the dataset mimics the imbalanced nature of the vulnerability detection problem.
	
	Using this dataset of methods, we used the DeepWater Framework (DWF) \cite{DWF} to facilitate the model running and evaluation tasks. DWF is a server-client architecture written in Python, using the well-known libraries scikit-learn and Tensorflow. It provides multiple pre-configured models, hyper-tuning, and resampling. The various models that we used provided by DWF are the following:
	
	\begin{description}
		\item[Adaboost:] Adaptive Boosting, an ensemble technique typically used for binary classification, which as part of it's learning phase adjusts the weak learners to better work on data points misclassified by previous weak-learners
		\item[CDNNC:] Custom Deep Neural Network, a feed-forward neural network with early stopping and L2 regularization
		\item[Forest:] Random Forest, an ensemble technique that uses (ideally not too deep) decision trees as weak learners
		\item[KNN:] K-nearest neighbors, a non-parametric method that predicts a data point's class by observing and aggregating the n closest (defined by a previously decided distance metric) datapoints 
		\item[Logistic:] Logistic Regression, a traditional method for classification that fits the sigmoid function on the given data
		\item[SDNNC:] Standard Deep Neural Network, a simple feed-forward neural network that is trained for a number of epochs 
		\item[Tree:] Decision Tree, a non-parametric technique that predicts a datapoint's class by learning decision rules
	\end{description}
	As our choice of evaluation metric, we used the F1-score, a widely accepted metric in situations where accuracy does not suffice because of the imbalanced nature of the dataset. It is calculated as the harmonic mean of precision ($TP/((TP+FP))$) and recall ($TP / (TP + FN)$), where $TP$ stand for true positive, $FP$ for false positive and $FN$ for false negative.
	
	All of these methods were hyper-tuned by generating 100 instances of each model with different parameters, where the parameters were explored using grid-search. The evaluation was done in a 10-fold cross-validation manner, with some special modifications to battle the imbalance problem that occurs both in the train set and in real-life scenarios: we used up-sampling to get 50\% vulnerable and non-vulnerable entries for each fold. The results can be seen in Table \ref{table:results}, where the Random Guesser stands for the bare minimum baseline that for any input predicts true with 20\% probability (equal to the ratio of positive-negative samples). The F1-score is calculated with its formula by considering the positive-negative ratio (0.2) and the probability of predicting true (0.2). In the following two RQs, we aim to interpret our results and also provide some more insight into our approach.
	
	\begin{table}
		\centering
		\scriptsize
		\caption{Results (F1-score)}
		\begin{tabular}{|l||c|c|c|c|c|c|c||l|} 
			\hline
			\textbf{Random Guesser} & \multicolumn{8}{c|}{20}                                                                                                                                                                                                                                                                          \\ 
			\hline
			& \multicolumn{1}{l|}{\textbf{Adaboost}} & \multicolumn{1}{l|}{\textbf{CDNNC}} & \multicolumn{1}{l|}{\textbf{Forest}} & \multicolumn{1}{l|}{\textbf{KNN}} & \multicolumn{1}{l|}{\textbf{Logistic}} & \multicolumn{1}{l|}{\textbf{SDNNC}} & \multicolumn{1}{l||}{\textbf{Tree}} & \textbf{Average}  \\ 
			\hline
			\textbf{Metrics}        & 36.99                                  & 31.84                               & 37.96                                & 38.32                             & 19.25                                  & 29.31                               & 36.80                               & 32.92             \\ 
			\hline
			\textbf{Simple }        & 40.62                                  & 40.33                               & 41.73                                & \textbf{47.46}                    & 29.82                                  & \textbf{44.33}                      & 40.20                               & 40.64             \\ 
			\hline
			\textbf{Change Tree}    & \textbf{41.37}                         & \textbf{44.38}                      & \textbf{43.38}                       & 43.28                             & \textbf{38.06}                         & 44.28                               & \textbf{41.69}                      & \textbf{42.34}    \\
			\hline
		\end{tabular}
		\label{table:results}
	\end{table}
	
	\subsection{RQ1: Can a vulnerability introducing database generated from a vulnerability fixing commit database be used for vulnerability prediction?}
	In this RQ, we reason about the usefulness of an automatically generated, vulnerability-related dataset. In more detail, it is important for such a dataset to be structured in a way that it can be easily used for downstream machine learning tasks. In our work, we focus on the problem of just-in-time vulnerability prediction, that is, given an instance of source code change by the before and after states, we predict its likeliness to be a change that leads to a vulnerable state of the software. We found that the dataset described in Section \ref{section:vuln_db} is suitable for such tasks, as it contains pairs of fixing and introducing commits and as such, it is straightforward to extract the before and after states. The database has a reasonable size and is filtered out of many false positives as a result of the filtering phase in our VIC dataset generator algorithm.
	
	After training multiple ML models and simulating their prediction power on unseen examples through 10-fold cross-validation we could achieve an F1-score averaging 45\% (see Table \ref{table:results}), which even though is far from optimal, it is substantially better than random guessing, and could be a good indicator for vulnerability prediction in a just-in-time manner. As a conclusion, we can say that we can generate accurate vulnerability introducing commit datasets from existing vulnerability fixing datasets in an automated way, that can be effectively used for training just-in-time vulnerability prediction ML models.
	
	\subsection{RQ2: How effective are Code Change Trees in representing source code changes? }
	Answering this RQ is based on the results shown in Table \ref{table:results} and the observation of average tree size changes between the simple change representation and Code Change Tree representation. 
	
	Firstly, the advantage of AST based representation forms over metrics based is evident as there is at least ~8\% increase in average F1-score. This finding further supports that AST based representations capture (most likely structural) information uncaught by metrics. A not that substantial, but still noticeable difference can be seen between the two AST based approaches, as Code Change Tree performs better by nearly 2\%.
	
	Another advantage that Code Change Tree has over the simple change representation is the tree size. We recorded the number of nodes in the trees for both representation forms for all the 59.340 functions that were considered as part of the change representations. In the case of the Code Change Tree, the average node number count was 51, while using the AST's in the simple code change representation it was 174. This is a reduction in size by more than 70\%, which means that Code Change Tree effectively reduces the size of representing change between source code states while still improving on the predictive power. To our understanding this is possible because Code Change Tree discards paths that are unchanged between the two states and consequently are irrelevant to the change.
	
	To summarize we can conclude that Code Change Trees perform substantially better than the metrics-based approach and marginally better than simply embedding and concatenating before and after states while reducing the tree size considerably.
	
	\section{Threats to validity}
	There are several issues that pose threats to the validity of the presented work.
	We use an automatically generated dataset for training just-in-time vulnerability prediction models and to evaluate the proposed Code Change Tree representation.
	Even if the original dataset of vulnerability fixing commits is fully validated, our method might introduce false entries to the generated dataset (i.e., code changes that are not introducing vulnerabilities).
	As this would hinder the conclusions of our study, we tried to mitigate the issue by performing a manual evaluation of the generated vulnerability introducing dataset on a random sample.
	We found that all the commits marked by our extraction method are linked to code changes introducing vulnerable functionality, therefore, the risk of relying on noisy data for evaluation is low.
	
	We compared our Code Change Tree based representation with naive approaches only (i.e., static source code metrics, token-based embedding).
	Therefore, we cannot state anything about its performance compared to other code change representations, like Commit2Vec \cite{commit2vec}.
	Even though it was out of the scope of this paper, we tried to use Commit2Vec but we were not able to find a publicly available implementation.
	However, an extensive comparative study of the just-in-time vulnerability detection capabilities of the various code change representations is in our future plans.
	
	We presented empirical results on vulnerability detection in Java systems, therefore our method to represent code changes and the prediction models might not generalize.
	However, as the method relies on the AST representation of the code in the before and after change states, it is easy to implement to other languages.
	If one could acquire a good quality vulnerability fixing dataset for a certain language, our method can be adapted.
	Nonetheless, a replication of the presented study might be desirable for other languages as well to increase the confidence of the generalizability of the approach.
	
	\section{Future work}
	As our work is a preliminary research on the usage of Code Change Trees related to just-in-time vulnerability detection and our results are promising, there are many possibilities for improvement. Mainly, we use a tree structure but lose a lot of its representative power by flattening. We plan on trying different ways of leveraging the information inherent to the AST derived tree structure in Code Change Tree, such as using graph neural networks \cite{gan}. 
	
	Also, it would be beneficial to compare our method to other similar ones, we also plan to extend the baselines with more complicated works than the ones we explored in this work.
	
	\section{Acknowledgement}
	The research was supported by the European Union project RRF-2.3.1-21-2022-00004 within the framework of the Artificial Intelligence National Laboratory and by project TKP2021-NVA-09.
	Project no. TKP2021-NVA-09 has been implemented with the support provided by the Ministry of Innovation and Technology of Hungary from the National Research, Development and Innovation Fund, financed under the TKP2021-NVA funding scheme.
	Furthermore, the research was partly financed by the EU-funded project AssureMOSS (Grant no. 952647).

	\bibliographystyle{splncs04}
	\bibliography{refs}

\end{document}